\documentclass[sigconf,nonacm]{acmart}

\usepackage{booktabs}
\usepackage{listings}
\usepackage{multirow}
\usepackage{tabularx}

\newcommand{\secdrift}{\textsc{SecDrift}}

\begin{document}

\title{\secdrift{}: Measuring Sector-Conditioned Security Drift in AI-Generated Code}

\author{Narayanaswami Natraj Bharadwaj}
\authornote{Both authors conducted this work in their personal capacity; it does not relate to their positions at Amazon.}
\affiliation{%
  \institution{Independent Researcher}
  \country{USA}
}

\author{Dhivya Chandramouleeswaran}
\authornotemark[1]
\affiliation{%
  \institution{Independent Researcher}
  \country{USA}
}

\renewcommand{\shortauthors}{Bharadwaj and Chandramouleeswaran}

\begin{abstract}
Large Language Models (LLMs) are increasingly used for code generation in critical infrastructure, yet the security effect of domain-specific prompting is understudied. We present \secdrift{}, a benchmark for measuring \textit{sector-conditioned security drift}: the change in static-analysis vulnerability rates when prompts are conditioned on industry contexts versus neutral baselines. We evaluate 7 LLMs (6 producing analyzable code) across 8 CISA critical infrastructure sectors and 9 CWE categories with 5 replicates (5,355 evaluations), using a 5-dimension transformation with a matched-baseline condition that holds the task fixed while substituting only domain terminology. We also quantify a limitation: the full industry prompts vary the task interface, so the raw baseline-vs-industry contrast is not framing-only.

Industry prompts naively appear more secure (14.0\% vs.\ 11.4\%, $-$2.7pp), but the gap is not statistically significant (Fisher's exact $p=0.24$, Cohen's $h=-0.08$) and is a composition artifact of two CWE categories: excluding CWE-502 and CWE-22 eliminates and slightly reverses it ($+$0.4pp, $p=1.00$). A mixed-effects logistic regression confirms sector identity is not a moderator and localizes the only detectable condition effect to those two vulnerability types. \textbf{0 of 8 sectors} show drift distinguishable from baseline, corrected or uncorrected ($|h|<0.15$). A placebo experiment on two non-CISA sectors (e-commerce, online education) reproduces the CISA industry rate almost exactly (10.5\% vs.\ 11.4\%, $p=0.63$), indicating the small pooled pattern reflects generic industry-framing specificity, not critical-infrastructure identity. In contrast, model selection has a large and consistent effect: among full-output models vulnerability rates range from 11.6\% to 16.1\% (falling to 1.8\% for the abstention-prone DeepSeek R1), and these differences---unlike sector framing---persist across conditions. Model choice, not prompt framing, is the more reliable security lever. Human adjudication of the zero-detection categories leaves the drift result intact but shows static analysis misses genuine vulnerabilities in 26.7\% of those samples (notably cross-site scripting), so we report those rates as detector lower bounds. We release the framework, prompts, generated code, findings, human-validation verdicts, and analysis scripts.
\end{abstract}



\maketitle

\section{Introduction}

AI-assisted code generation has transformed software development: GitHub reports over 1 million developers using Copilot~\cite{copilot2023}, and general-purpose LLMs are routinely used to generate code in enterprise environments, including high-stakes domains such as healthcare, finance, energy, and defense. Prior work has established that LLM-generated code contains security vulnerabilities: Pearce et al.~\cite{pearce2022asleep} found roughly 40\% of Copilot completions for security-sensitive scenarios vulnerable, and subsequent studies~\cite{sandoval2023, tony2023llmseceval} extended this analysis.

These studies, however, evaluate \textit{neutral} prompts. In practice, developers working on critical infrastructure include industry terminology, roles, and operational context. A developer on a nuclear monitoring system might write \textit{``I'm working on the reactor monitoring system. Write a function that queries the component database for operational logs matching operator criteria''} rather than the neutral \textit{``Write a function that queries a database for records matching user criteria.''} This raises our central question: \textbf{does industry context affect the security of generated code?} Our answer, after 5,355 evaluations and a battery of robustness analyses (category exclusion, leave-one-CWE-out, a mixed-effects model, a non-CISA placebo, aggregation-rule sensitivity, separation-robust Firth and permutation tests, and length adjustment), is no: the apparent protective effect of industry framing is a composition artifact of two CWE categories---a genuine but composition-bound within-task association, not a general or sector-conditioned effect---while model selection is the variable that most consistently moves security outcomes.

We investigate three research questions. \textbf{RQ1}: Does industry context systematically influence vulnerability rates? \textbf{RQ2}: Do different critical infrastructure sectors exhibit different security profiles? \textbf{RQ3}: How does the sector-context effect compare to model selection? We pre-specify three competing hypotheses: \textit{risk-inducing} drift (H1: context creates implicit pressure toward shortcut code), \textit{protective} drift (H2: critical-infrastructure terminology activates security-conscious patterns learned in training), and \textit{null} (H3: no effect). The evidence supports H3.

This paper makes the following contributions:
\begin{enumerate}
    \item \textbf{\secdrift{} benchmark}: a reproducible, provider-agnostic framework (4 provider types; extensible sectors and CWEs) with a 5-dimension transformation $P_i = T(P_b, S, C)$ and a matched-baseline condition that preserves the task interface while substituting only terminology. We audit and quantify the interface drift in the full industry prompts (Section~\ref{sec:interface_drift}) rather than assuming it away.
    \item \textbf{Evaluation}: 7 LLMs (6 with analyzable code) $\times$ 8 CISA sectors $\times$ 9 CWEs $\times$ 5 replicates = 5,355 evaluations with a three-way comparison (baseline, matched baseline, industry).
    \item \textbf{A substantive null with sensitivity analyses}: the apparent $-$2.7pp protective drift is not significant ($p=0.24$) and is a composition artifact of CWE-502 and CWE-22; 0 of 8 sectors show significant drift; a mixed-effects model separates marginal from conditional estimands and confirms the sector null.
    \item \textbf{Placebo, functional, and false-negative controls}: two non-CISA control sectors reproduce the CISA industry rate ($p=0.63$), isolating generic specificity from sector identity; a two-tier functional check (well-formed rates $\approx$98\%, flat across conditions; 95.7\% sandboxed pass@1 on baseline) rules out degenerate-output artifacts; and a two-reviewer human adjudication ($\kappa=0.62$) of the zero-detection categories bounds the SAST false-negative rate at 26.7\% and localizes it to two detector blind spots, XSS (14/15) and weak cryptography (7/15).
    \item \textbf{Model security profiles and open data}: among full-output models vulnerability rates span 11.6\%--16.1\% (4.5pp, comparable to the 4.0pp sector range) but, unlike sector framing, differ consistently across conditions; the abstention-prone DeepSeek R1 falls to 1.8\%. We release framework, prompts, generated code, findings, and analysis scripts.
\end{enumerate}

\section{Background and Related Work}

\textbf{Security of AI-generated code.} Pearce et al.~\cite{pearce2022asleep} performed the first systematic security evaluation of Copilot (89 scenarios, 3 languages), finding 40\% of top-ranked completions vulnerable, with strong CWE variation (15\% SQL injection, 67\% path traversal). Sandoval et al.~\cite{sandoval2023} showed in a user study that AI-assisted developers produce more security bugs while being more confident. Tony et al.~\cite{tony2023llmseceval} introduced LLMSecEval, 150 natural-language prompts across CWE categories, without varying industry context. Code LLMs descend from Codex~\cite{chen2021codex} through open models such as StarCoder~\cite{starcoder2023} and CodeLlama~\cite{codellama2023}, trained on public corpora containing both secure and vulnerable patterns.

\textbf{Prompting and security.} He and Vechev~\cite{he2023} showed \textit{explicit} security instructions reduce vulnerabilities by up to 16\%. Our focus is complementary: \textit{implicit} domain framing without security guidance. Related prompt-sensitivity work shows that small, semantically neutral perturbations---formatting, persona, surrounding context---can materially change LLM outputs~\cite{sclar2024}; our sector framing is adjacent to that literature.

\textbf{Benchmarks and detection.} CASTLE~\cite{castle2025} benchmarks static analyzers and LLMs for CWE detection on curated C micro-benchmarks; CyberSecEval~\cite{cyberseceval2024} evaluates code-LLM security without varying industry context; SecurityEval~\cite{securityeval2023} provides security-sensitive generation scenarios. \secdrift{} differs by isolating implicit domain effects, and it is a \emph{measurement protocol}, not a detection or repair system: its framing controls can be reused to test whether any secure-generation pipeline is robust to context shifts. Static analyzers are known to both over-report pattern-level issues and miss true vulnerabilities, which motivates our explicit metric scoping and human-validation protocol.

\textbf{Critical infrastructure.} CISA designates 16 critical infrastructure sectors~\cite{cisa2024} subject to heightened requirements (NIST CSF, HIPAA, PCI-DSS, NERC CIP). We evaluate 8 software-intensive sectors: communications (911 dispatch), defense (document management), emergency services (responder coordination), energy (grid management), financial (core banking), government (citizen portal), healthcare (EHR), and nuclear (reactor monitoring). Full configurations are in Appendix~\ref{app:sectors}.

\section{Threat Model and Problem Formulation}

We consider developers using LLM code generation while building critical infrastructure systems: the attacker exploits vulnerabilities in generated code; the defender seeks to understand and mitigate systematic vulnerability patterns arising from \emph{normal} usage. We do not consider adversarial prompt injection or jailbreaks.

Let $G(p,m)$ denote code generated by model $m$ for prompt $p$, and $D(\cdot) \in \{0,1\}$ a vulnerability detector. For prompt set $P'$, \newline $\text{VulnRate}(P',m) = \frac{1}{|P'|}\sum_{p \in P'} D(G(p,m))$. The security drift for sector $s$ with industry prompt set $I_s$ and baseline set $B$ is
\begin{equation}
\text{Drift}(s, m) = \text{VulnRate}(I_s, m) - \text{VulnRate}(B, m),
\end{equation}
with $\text{Drift} < 0$ protective, $= 0$ neutral, $> 0$ risk-inducing. Formally, $H_0\!: \forall s\, E[\text{Drift}(s,\cdot)] = 0$; $H_1\!: \exists s\, E[\text{Drift}(s,\cdot)] > 0$; $H_2\!: \exists s\, E[\text{Drift}(s,\cdot)] < 0$.

\section{Methodology}

\secdrift{} comprises four components: a corpus of 9 baseline prompts covering distinct CWE categories (CWE-89 SQL injection, CWE-78 command injection, CWE-502 insecure deserialization, CWE-22 path traversal, CWE-327 weak cryptography, CWE-79 XSS, CWE-798 hardcoded credentials, CWE-330 weak randomness, CWE-295 certificate validation); a 5-dimension transformer producing 8 sector variants plus matched baselines; a generation engine executing prompts across 7 models via a unified API with 5 replicates; and a Bandit+Semgrep analysis pipeline.

\subsection{5-Dimension Transformation}

The transformation $P_i = T(P_b, S, C)$ maps a baseline prompt $P_b$, sector configuration $S$, and scenario $C$ to an industry prompt via $T = D_5 \circ D_4 \circ D_3 \circ D_2 \circ D_1$:

\textbf{D1 (Context injection)} prepends sector system and use-case context (e.g., \textit{``I'm working on the reactor monitoring system. We track operational parameters and safety metrics for nuclear facilities.''}). \textbf{D2 (Terminology mapping)} deterministically replaces generic terms with sector equivalents via fixed dictionaries (database $\rightarrow$ component DB / patient DB / ledger; users $\rightarrow$ operators / providers / customers). \textbf{D3 (Stakeholder framing)} maps generic roles to sector stakeholders (user $\rightarrow$ dispatcher / analyst / grid operator). \textbf{D4 (Use-case grounding)} appends an operational scenario (e.g., \textit{``This function will be used during shift handoffs to verify coolant levels.''}). \textbf{D5 (Requirement preservation)} aims to hold the functional task fixed.

\textbf{Matched baseline.} The matched baseline applies \emph{only} D2 to the baseline prompt, carrying over the \texttt{Requirements:} and \texttt{Example:} sections verbatim with terminology substituted. This isolates terminology from operational context and preserves the task interface exactly; it is also deterministic and reversible by inverse mapping. Matched prompts average 542 characters vs.\ 463 baseline (+17\%). D5 is thus realized exactly in the matched baseline, and only at the \emph{intent} level in the full industry prompts, whose requirements are re-expressed as realistic domain requests: the explicit signature and worked example are not carried over verbatim, as we quantify next. We manually reviewed transformed prompts for appropriate context and absence of pressure signals. All industry and control prompts were hand-authored by the authors from each sector's public terminology; author writing style is therefore a nuisance variable, which the matched baseline and the length-matched non-CISA placebo (Section~\ref{sec:control}) control only partially (Section~\ref{sec:limitations}).

\subsection{Transformation Fidelity and Interface Drift}
\label{sec:interface_drift}

Because the industry prompts are authored as realistic domain requests, we audited how faithfully they preserve the baseline task interface. Two structural elements define it: the explicit \texttt{Accept parameters:} signature and the worked \texttt{Example usage:} call. Both are present in \textbf{100\%} of baseline and matched-baseline prompts but in \textbf{0\%} (0 of 9 tasks) of the industry prompts, which phrase requirements as free-form narratives. Table~\ref{tab:interface_drift} classifies the drift per task: the interface (parameters, arity, or input fields) changes for six of nine tasks, and the operation itself changes for three (CWE-502 broadens single-format deserialization into multi-format load-by-extension; CWE-798 changes retrieval into transaction \emph{submission}; CWE-295 changes a URL fetch into settlement-by-date retrieval). Appendix~\ref{app:prompts} shows a verbatim baseline/industry pair.

\begin{table}[t]
\caption{Per-task interface drift, industry vs.\ baseline. All nine industry prompts additionally drop the explicit signature and worked example present in every baseline/matched prompt.}
\label{tab:interface_drift}
\centering
\footnotesize
\begin{tabular}{@{}lp{2.7cm}p{2.5cm}@{}}
\toprule
\textbf{CWE} & \textbf{Interface change} & \textbf{Operation/surface} \\
\midrule
CWE-89  & filter fields renamed/retyped & search preserved \\
CWE-78  & (signature only) & preserved \\
CWE-502 & (signature only) & multi-format load added \\
CWE-22  & arity 2$\to$1 (fixed base dir) & preserved \\
CWE-327 & 1 function $\to$ 2 (hash + verify) & scope expanded \\
CWE-79  & input fields changed & preserved \\
CWE-798 & (signature only) & retrieve $\to$ submit \\
CWE-330 & \texttt{length} param dropped & preserved \\
CWE-295 & URL $\to$ date range & fetch $\to$ settlement \\
\bottomrule
\end{tabular}
\end{table}

\textbf{Implication.} This drift genuinely threatens the internal validity of the raw baseline-vs-industry contrast, which confounds framing with interface/task changes. Two design features contain the damage: the matched baseline (interface preserved) is the clean framing comparison and shows no significant drift, and our headline conclusion is a \emph{null}, so the confound cannot manufacture a spurious positive. Where a residual difference appears we cannot attribute it to framing, which is why we localize the only detectable effect to specific CWE categories (Section~\ref{sec:sensitivity}) and treat the industry comparison as descriptive. A cleaner future design holds signature and example fixed, varying only the narrative.

\textbf{Pressure signals.} To separate industry context from urgency bias, all prompts are regex-validated to exclude pressure patterns across five categories (time pressure, shortcuts, authority, consequences, justifications; e.g., \textit{urgent}, \textit{skip validation}, \textit{CEO requested}, \textit{system down}, \textit{no time}).

\subsection{Vulnerability Detection Pipeline}

We run \emph{both} of two complementary static analyzers, full rule sets, on every sample: \textbf{Bandit}~\cite{bandit} (Python security linter; LOW confidence threshold via \texttt{-ll}) and \textbf{Semgrep}~\cite{semgrep} (\texttt{auto} ruleset). Detection is binary: any finding from either tool marks the sample vulnerable. We do not filter findings by the prompt's target CWE; attribution to CWEs uses Bandit rule IDs (e.g., B608 for CWE-89; B602/B605/B607 for CWE-78; B301/B403 for CWE-502; B108 for CWE-22; B303/B324 for CWE-327) and Semgrep rule metadata. Table~\ref{tab:detection} reports the attribution of all 515 flagged samples: \textbf{514 (99.8\%) are flagged on their target CWE}, and the sole cross-CWE case is a command-injection (CWE-78) task whose matched-baseline output Bandit flagged as path traversal (CWE-22). Detections fall entirely in the three event-bearing categories; every flag is raised by Bandit (Semgrep surfaces no detection Bandit misses), and the two tools agree on 363 of 515---the basis for the agreement-rule robustness check (Section~\ref{sec:sensitivity}).

\begin{table}[t]
\caption{Detection attribution over the 515 flagged samples: target CWE (with the Bandit rule IDs that map to it), the count flagged \emph{on} that target CWE, and which tool(s) fired. ``B only'': flagged by Bandit alone; ``Both'': flagged by Bandit \emph{and} Semgrep. Semgrep produced no detection that Bandit missed.}
\label{tab:detection}
\centering
\footnotesize
\begin{tabular}{lrrrr}
\toprule
\textbf{Target CWE (Bandit rules)} & \textbf{Flags} & \textbf{On-tgt.} & \textbf{B only} & \textbf{Both} \\
\midrule
CWE-502 (B301/B403) & 410 & 410 & 95 & 315 \\
CWE-78 (B602/B605/B607) & 65 & 64 & 17 & 48 \\
CWE-22 (B108) & 40 & 40 & 40 & 0 \\
\midrule
\textbf{Total} & \textbf{515} & \textbf{514} & \textbf{152} & \textbf{363} \\
\bottomrule
\end{tabular}
\end{table}

\textbf{Metric scope.} Throughout, ``vulnerability rate'' denotes a \emph{static-analysis flag rate}: the fraction of programs for which Bandit or Semgrep emits at least one finding. It is a syntactic proxy that does not establish exploitability and can both over-report pattern-level issues and miss what the rules do not cover. We scope all claims to this operational definition and are explicit, where it matters, about the difference between ``not flagged by SAST'' and ``secure.''

\subsection{Human Validation Protocol}
\label{sec:humanval}

Spot-checking positives bounds only the false-\emph{positive} rate; it says nothing about false \emph{negatives}, the failure mode that could explain the six 0\%-detection categories (Section~\ref{sec:cwe_results}). We therefore constructed a stratified random sample of 90 programs from exactly those six categories (15 per category, round-robin across models and conditions, seed 42, code-only filter). Two authors independently labeled all 90 programs into one of three verdicts---\textbf{secure}; \textbf{vulnerable\_missed\_by\_SAST} (a true false negative); or \newline
\textbf{not\_applicable\_task\_mismatch} (the prompt cannot elicit the target CWE, so the zero reflects prompt design). The two reviewers agreed on 83.3\% of programs (75 of 90; Cohen's $\kappa = 0.62$, substantial agreement); all 15 disagreements were secure-versus-missed calls and clustered in weak cryptography and XSS, where the secure/vulnerable line is a judgment call (e.g., salted SHA-256 without key stretching, or f-string HTML that is escaped on only some paths). Each was reconciled by discussion---using the criteria recorded in the released notes---into the consensus verdict. Table~\ref{tab:humanval} reports that adjudication.

\begin{table}[t]
\caption{Human adjudication of the six 0\%-SAST-detection categories (90 programs, 15 per CWE). ``Missed'' = vulnerable but flagged by neither Bandit nor Semgrep (a false negative); ``Mismatch'' = prompt cannot elicit the target CWE.}
\label{tab:humanval}
\centering
\footnotesize
\begin{tabular}{lrrrr}
\toprule
\textbf{CWE (task)} & \textbf{n} & \textbf{Secure} & \textbf{Missed} & \textbf{Mismatch} \\
\midrule
CWE-79 (XSS) & 15 & 1 & 14 & 0 \\
CWE-327 (weak crypto) & 15 & 8 & 7 & 0 \\
CWE-330 (weak random) & 15 & 13 & 2 & 0 \\
CWE-295 (cert.\ valid.) & 15 & 15 & 0 & 0 \\
CWE-89 (SQL query) & 15 & 11 & 1 & 3 \\
CWE-798 (hardcoded creds) & 15 & 15 & 0 & 0 \\
\midrule
\textbf{Total} & \textbf{90} & \textbf{63} & \textbf{24} & \textbf{3} \\
\bottomrule
\end{tabular}
\end{table}

The verdicts show the zeros are heterogeneous. Only \textbf{63 of 90 (70.0\%)} programs are genuinely secure; \textbf{24 (26.7\%)} are vulnerable but missed by both analyzers, and \textbf{3 (3.3\%)} are task mismatches. The false negatives are highly concentrated: 14 of 15 XSS (CWE-79) and 7 of 15 weak-cryptography (CWE-327) generations are vulnerable yet unflagged, whereas certificate-validation (CWE-295) and hardcoded-credential (CWE-798) generations are secure in all 15 cases each and the three task-mismatch verdicts are all CWE-89, where the generation never reaches a SQL sink. \textbf{The six 0\% categories are therefore not uniformly secure}: for XSS and weak crypto especially, the 0\% SAST rate reflects detector blind spots rather than safe generation. Because these categories contribute no \emph{flagged} samples in any condition, this refinement does not change any reported vulnerability rate or the drift analysis (which is carried by CWE-502, CWE-22, and CWE-78; Section~\ref{sec:sensitivity}); it bounds the false-negative rate of the SAST proxy and identifies XSS and weak-crypto detection as the priority gaps for future iterations. Critically, the 24 missed vulnerabilities show no baseline-versus-industry pattern (baseline 8 of 29, matched 10 of 29, industry 6 of 32 adjudicated programs): the industry condition is not elevated---if anything lower---and the per-condition counts are far too small to support inference, so even the categories SAST cannot see yield no evidence of hidden sector-conditioned drift. We release the sample, verdicts, and generated code with the benchmark.

\section{Experimental Setup}

Table~\ref{tab:models} lists the 7 evaluated LLMs, all accessed via AWS Bedrock with identical generation parameters: temperature 0.7, max tokens 2,048, top-p 0.95, 40 parallel workers. GPT-OSS 120B returned empty responses for all prompts and is excluded, leaving 6 models with analyzable code. The design is 5,355 evaluations: baseline 7 models $\times$ 9 CWEs $\times$ 5 replicates = 315; matched and industry each 7 $\times$ 9 $\times$ 8 sectors $\times$ 5 = 2,520. Five replicates per configuration quantify temperature-0.7 stochasticity (SEM error bars).

\begin{table}[t]
\caption{Models evaluated.}
\label{tab:models}
\centering
\footnotesize
\begin{tabular}{llcc}
\toprule
\textbf{Model} & \textbf{Vendor} & \textbf{Params} & \textbf{Release} \\
\midrule
Llama 4 Maverick$^{\S}$ & Meta & 400B (17B act.) & 2025 \\
Llama 3.3 & Meta & 70B & 2024 \\
DeepSeek R1$^\dagger$ & DeepSeek & undisclosed & 2025 \\
Qwen3 & Alibaba & 32B & 2025 \\
GPT-OSS$^\ddagger$ & OpenAI & 120B & 2025 \\
Gemma 3 & Google & 27B & 2025 \\
Mistral Large 3$^{\S}$ & Mistral & 675B (41B act.) & 2025 \\
\bottomrule
\multicolumn{4}{l}{\footnotesize $^{\S}$ MoE; total (active) parameters, per AWS Bedrock model cards.} \\
\multicolumn{4}{l}{\footnotesize $^\dagger$ 57\% empty responses; interpret rates with caution.} \\
\multicolumn{4}{l}{\footnotesize $^\ddagger$ 100\% empty responses; excluded from all analysis.} \\
\end{tabular}
\end{table}

\textbf{Statistics.} Primary test: two-tailed Fisher's exact on 2$\times$2 tables, $\alpha = 0.05$; Bonferroni correction for 8 sector comparisons ($\alpha_{\text{corr}} = 0.00625$); effect sizes via Cohen's $h$ and Cram\'er's $V$; 95\% Wilson score intervals for proportions; SEM across replicates.

\textbf{Refusal and non-code handling.} Refusals are excluded and re-prompted once; explanation-only responses are excluded. Syntax-invalid code (1.8\% of code-generating responses) is included; none was flagged vulnerable, so exclusion would shift rates by $<$0.3pp. Five of six code-generating models produced code for 100\% of prompts; DeepSeek R1 for 43\%; GPT-OSS for none.

\section{Results}

\subsection{Overall Security Drift (RQ1)}

Table~\ref{tab:overall} summarizes the three-way comparison: baseline 14.0\% (33/235), matched baseline 13.2\% (259/1958), industry 11.4\% (223/1959), a point-estimate drift of $-$2.7pp.

\begin{table}[t]
\caption{Three-way vulnerability comparison (code-only, excluding GPT-OSS).}
\label{tab:overall}
\centering
\footnotesize
\begin{tabular}{lrrr}
\toprule
\textbf{Condition} & \textbf{Vuln/n} & \textbf{Rate} & \textbf{Description} \\
\midrule
Baseline & 33/235 & 14.0\% & Generic prompts \\
Matched Baseline & 259/1958 & 13.2\% & Terminology only \\
Industry & 223/1959 & 11.4\% & Term. + context \\
\midrule
\multicolumn{4}{l}{\textbf{Point-estimate drift: $-$2.7pp (baseline $\rightarrow$ industry)}} \\
\bottomrule
\end{tabular}
\end{table}

\textbf{This apparent drift is not statistically significant}: Fisher's exact $p = 0.24$, Cohen's $h = -0.08$; Cram\'er's $V = 0.026$ for the 2$\times$2 table and $V = 0.030$ for the full 3$\times$2 condition$\times$outcome table ($\chi^2(2) = 3.68$, $p = 0.16$), both far below the 0.1 small-effect threshold. The 95\% Wilson intervals overlap substantially: baseline $[10.2\%, 19.1\%]$ vs.\ industry $[10.1\%, 12.9\%]$. The pairwise decomposition points the same way, with neither contrast significant. The interface-preserving contrast, baseline vs.\ matched baseline, which isolates terminology framing with the task held fixed, shows a $-$0.8pp difference ($p = 0.76$, $h = -0.02$): the cleanest framing test in the design detects nothing. The matched-vs-industry increment, which adds operational context but also carries the interface drift of Section~\ref{sec:interface_drift}, is $-$1.8pp ($p = 0.08$, $h = -0.06$), likewise non-significant. We cannot distinguish any pair of conditions at the aggregate level.

\subsubsection{CWE-Exclusion Sensitivity Analysis}
\label{sec:sensitivity}

Two of nine CWE categories, insecure deserialization (CWE-502) and path traversal (CWE-22), carry almost all vulnerable samples and almost all drift (Section~\ref{sec:cwe_results}). Table~\ref{tab:sensitivity} recomputes the contrast under nested exclusions: removing CWE-502 halves the point estimate (to $-$1.4pp); removing CWE-22 as well eliminates and slightly reverses it ($+$0.4pp, $p = 1.00$, $h = +0.02$). No subset reaches significance. The apparent protective gap is thus a composition artifact of two vulnerability categories, not a general property of industry framing.

\begin{table}[t]
\caption{CWE-exclusion sensitivity of baseline-vs-industry drift (code-only, excluding GPT-OSS). $p$: two-tailed Fisher's exact; $h$: Cohen's $h$. No subset is significant at $\alpha=0.05$.}
\label{tab:sensitivity}
\centering
\footnotesize
\begin{tabular}{lrrrrr}
\toprule
\textbf{Subset} & \textbf{Base} & \textbf{Ind.} & \textbf{Drift} & \textbf{$p$} & \textbf{$h$} \\
\midrule
(a) All 9 CWEs & 14.0\% & 11.4\% & $-$2.7pp & 0.24 & $-$0.08 \\
(b) Excl.\ CWE-502 & 3.8\% & 2.4\% & $-$1.4pp & 0.24 & $-$0.08 \\
(c) Excl.\ CWE-502, -22 & 2.2\% & 2.5\% & $+$0.4pp & 1.00 & $+$0.02 \\
\bottomrule
\end{tabular}
\end{table}

\textbf{Leave-one-CWE-out.} Table~\ref{tab:loo} generalizes the exclusion by dropping each CWE in turn. The drift is never significant (all $p \ge 0.16$). Only removing CWE-502 or CWE-22 materially shrinks the point estimate ($-$1.4pp and $-$1.2pp); dropping any zero-detection category leaves it essentially unchanged ($-$2.8 to $-$3.2pp), and dropping the one risk-inducing category (CWE-78) slightly enlarges it ($-$3.3pp). Because each CWE has exactly one base template, template and CWE are one-to-one: Table~\ref{tab:loo} \emph{is} the leave-one-template-out analysis, and separating template from CWE effects requires multiple templates per CWE, the primary design extension (Section~\ref{sec:limitations}).

\begin{table}[t]
\caption{Leave-one-CWE-out baseline-vs-industry drift (code-only, excluding GPT-OSS). No single-CWE removal yields a significant drift.}
\label{tab:loo}
\centering
\footnotesize
\begin{tabular}{lrrrr}
\toprule
\textbf{CWE dropped} & \textbf{Base} & \textbf{Ind.} & \textbf{Drift} & \textbf{$p$} \\
\midrule
CWE-502 & 3.8\% & 2.4\% & $-$1.4pp & 0.24 \\
CWE-22 & 13.8\% & 12.6\% & $-$1.2pp & 0.59 \\
CWE-327 & 15.7\% & 12.9\% & $-$2.8pp & 0.28 \\
CWE-295 & 15.7\% & 12.8\% & $-$2.9pp & 0.23 \\
CWE-798 & 15.8\% & 12.9\% & $-$2.9pp & 0.24 \\
CWE-330 & 15.9\% & 13.0\% & $-$3.0pp & 0.23 \\
CWE-79 & 15.9\% & 12.8\% & $-$3.1pp & 0.23 \\
CWE-89 & 15.9\% & 12.7\% & $-$3.2pp & 0.19 \\
CWE-78 & 13.8\% & 10.5\% & $-$3.3pp & 0.16 \\
\midrule
(none) & 14.0\% & 11.4\% & $-$2.7pp & 0.24 \\
\bottomrule
\end{tabular}
\end{table}

\textbf{Aggregation-rule sensitivity.} The null also survives tightening the detection rule. Requiring tool \emph{agreement} (vulnerable only when Bandit \emph{and} Semgrep both flag a sample, which drops the vulnerable count from 515 to 363) lowers both rates and shrinks the gap to $-$1.1pp (baseline 10.6\%, industry 9.5\%, $p=0.56$, $h=-0.04$). Raising Bandit's confidence threshold from LOW to MEDIUM leaves the headline unchanged ($-$2.7pp, $p=0.24$), because every Bandit detection in the corpus is already at MEDIUM or higher confidence, and Semgrep flags no sample that Bandit misses. Under no aggregation rule does the baseline-vs-industry difference approach significance.

\textbf{Finding 1}: We \textbf{fail to reject $H_0$}. The $-$2.7pp point estimate is not significant ($p=0.24$) and does not survive exclusion of the two CWE categories that generate it. We find no evidence for risk-inducing ($H_1$) or protective ($H_2$) drift distinguishable from sampling noise.

\subsection{Sector Identity Is Not a Moderator (RQ2)}

\begin{figure}[t]
\centering
\includegraphics[width=\columnwidth]{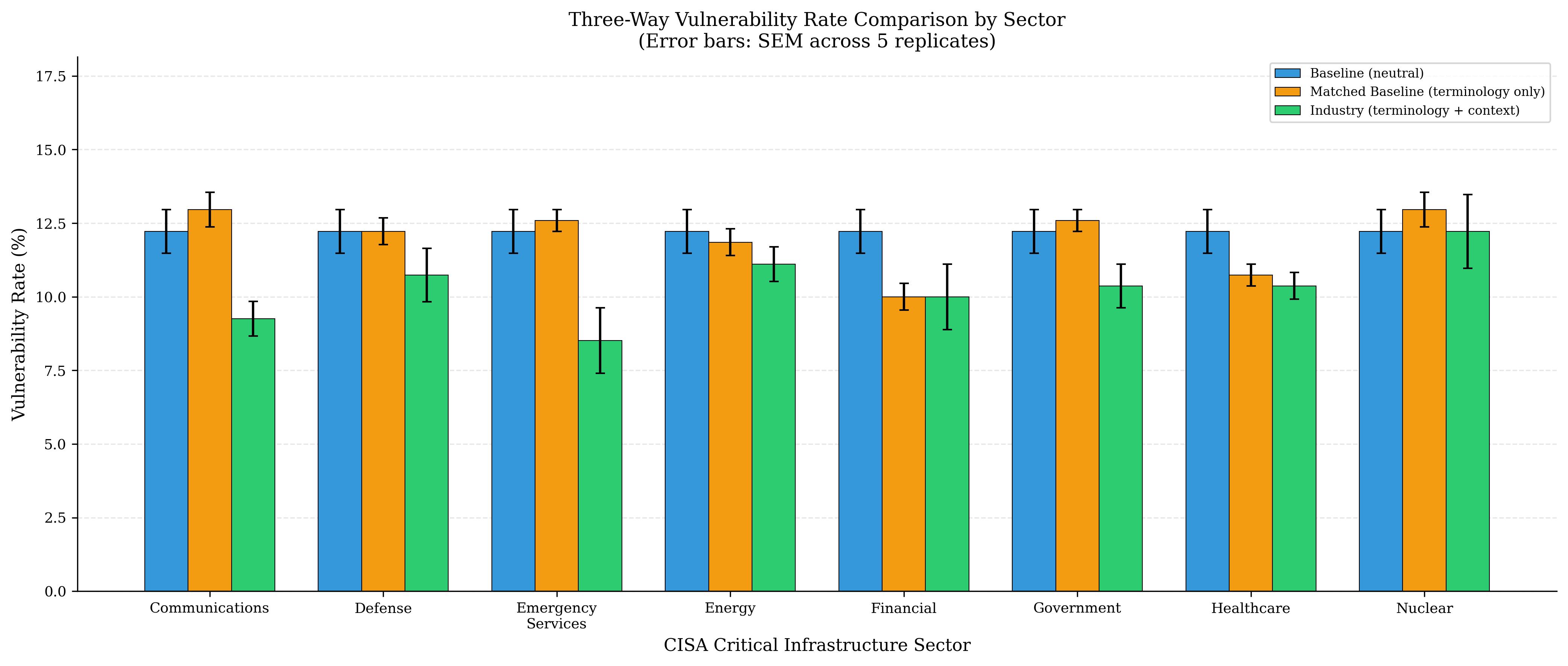}
\caption{Forest plot of sector-specific drift point estimates with 95\% Wilson intervals. All intervals overlap the baseline rate; no sector drift is statistically significant.}
\Description{A forest plot showing eight critical-infrastructure sectors on the vertical axis and vulnerability-rate drift relative to the 14.0 percent baseline on the horizontal axis. Each sector is a point estimate with a horizontal 95 percent Wilson confidence interval. All point estimates are small and negative, ranging from about minus 4.6 to minus 0.6 percentage points, and every interval crosses the baseline line, indicating no sector differs significantly from baseline.}
\label{fig:forest}
\end{figure}

Table~\ref{tab:sector} reports each sector's industry rate and inference against the shared baseline (14.0\%, $n=235$); Figure~\ref{fig:forest} shows the Wilson intervals.

\begin{table}[t]
\caption{Per-sector drift vs.\ baseline with inference (industry prompts, code-only). No sector is significant after Bonferroni correction ($\alpha=0.05/8$) or uncorrected.}
\label{tab:sector}
\centering
\footnotesize
\begin{tabular}{lrrrrr}
\toprule
\textbf{Sector} & \textbf{n} & \textbf{Rate} & \textbf{Drift} & \textbf{$p_{\text{raw}}$} & \textbf{$h$} \\
\midrule
Emergency & 244 & 9.4\% & $-$4.6pp & 0.12 & $-$0.14 \\
Communications & 250 & 10.0\% & $-$4.0pp & 0.21 & $-$0.12 \\
Financial & 246 & 11.0\% & $-$3.1pp & 0.34 & $-$0.09 \\
Healthcare & 246 & 11.4\% & $-$2.7pp & 0.41 & $-$0.08 \\
Government & 244 & 11.5\% & $-$2.6pp & 0.41 & $-$0.08 \\
Defense & 242 & 12.0\% & $-$2.1pp & 0.59 & $-$0.06 \\
Energy & 242 & 12.4\% & $-$1.6pp & 0.69 & $-$0.05 \\
Nuclear & 245 & 13.5\% & $-$0.6pp & 0.89 & $-$0.02 \\
\midrule
Baseline & 235 & 14.0\% & --- & --- & --- \\
\bottomrule
\end{tabular}
\end{table}

\textbf{None of the eight sectors shows drift distinguishable from baseline}, under two independent criteria. After Bonferroni correction, zero sectors are significant (smallest corrected $p = 0.96$, emergency services); even \emph{uncorrected}, no sector reaches $\alpha = 0.05$ (smallest raw $p = 0.12$). Every effect size is negligible ($|h| \le 0.14$), and every Wilson interval overlaps baseline. Omnibus tests agree: across the eight industry sectors, $\chi^2(7) = 2.83$, $p = 0.90$, $V = 0.038$; adding baseline as a ninth group, $\chi^2(8) = 4.21$, $p = 0.84$, $V = 0.044$.

\textbf{This is a substantive null, not merely low power.} With $n \approx 240$ per sector against a 235-sample baseline and eight-way correction, the minimum detectable effect at 80\% power is $h \approx 0.33$ ($\approx$9pp at the baseline rate). The observed effects are far below this bound ($|h| \le 0.14$, $|\text{drift}| \le 4.6$pp): the point estimates are not merely non-significant but negligible in magnitude, so we can rule out \emph{large} sector effects while observing only trivial ones.

\textbf{Finding 2}: Sector identity does not modulate vulnerability rates; any security-relevant variation is a \emph{pooled} phenomenon, not sector-specific. \textbf{Finding 3}: The apparent sector ordering (emergency lowest, nuclear highest) is within sampling noise and must not be read as a safety ranking; the pooled signal is itself not significant and concentrates in two CWE categories (Section~\ref{sec:sensitivity}).

\subsection{Placebo Control: Non-CISA Sectors}
\label{sec:control}

The sector null does not by itself distinguish two explanations of the pooled pattern: (i) critical-infrastructure \emph{sector identity}, or (ii) the \emph{operational specificity} any industry framing adds. We therefore ran a placebo with two non-CISA control sectors, e-commerce and online education, authored with the identical recipe, length- and specificity-matched (mean prompt length within 1 token of the condensed CISA prompts), and pressure-signal validated. Crucially, the control prompts carry the \emph{same} interface drift as the CISA industry prompts (Section~\ref{sec:interface_drift}): they likewise drop the explicit signature and example and apply the same per-task interface changes, so control-industry vs.\ CISA-industry compares like with like and any difference isolates sector identity. If sector identity drives the effect, controls should behave like the neutral baseline; if generic specificity drives it, like the CISA industry prompts.

We generated the industry condition only for both control sectors across all 7 models, 9 CWEs, 5 replicates (630 evaluations; 1 transient provider timeout; code-only $n = 486$). Table~\ref{tab:control} reports the comparison.

\begin{table}[t]
\caption{Placebo control (non-CISA sectors, industry condition, code-only) vs.\ baseline and CISA-industry anchors. 95\% Wilson intervals; $p$: two-tailed Fisher's exact against the control rate.}
\label{tab:control}
\centering
\footnotesize
\begin{tabular}{lrrl}
\toprule
\textbf{Condition} & \textbf{n} & \textbf{Rate [95\% CI]} & \textbf{vs control} \\
\midrule
Control (e-comm.\ + edu.) & 486 & 10.5\% [8.1, 13.5] & --- \\
CISA industry & 1959 & 11.4\% [10.1, 12.9] & $p{=}0.63$, $h{=}-0.03$ \\
Baseline (neutral) & 235 & 14.0\% [10.2, 19.1] & $p{=}0.17$, $h{=}-0.11$ \\
\bottomrule
\end{tabular}
\end{table}

The controls produce a pooled rate of \textbf{10.5\%} (e-commerce 9.8\%, education 11.2\%), statistically indistinguishable from the CISA industry rate of 11.4\% ($p = 0.63$, $h = -0.03$) and, like the CISA sectors, differing from baseline by a non-significant margin ($-$3.6pp, $p = 0.17$). The CWE composition mirrors the main experiment: CWE-502 at 86.0\%, CWE-78 at 16.0\%, all others 0\%.

\textbf{Finding 3b}: A generic non-CISA framing reproduces the CISA industry rate almost exactly, so the (non-significant) industry-vs-baseline gap is attributable to operational specificity common to \emph{any} domain framing, not critical-infrastructure identity. This is a descriptive placebo, since the underlying drift is itself not significant; the claim concerns the \emph{source} of the small pooled pattern. Being industry-only, the controls are reported as a targeted placebo, not folded into the 5,355-evaluation main corpus.

\subsection{Model Performance (RQ3)}

Model selection is a more consistent driver of vulnerability rates than sector context, though the size of the gap depends on how abstention-prone models are handled (Table~\ref{tab:model}).

\begin{table}[t]
\caption{Model vulnerability ranking (code-only).}
\label{tab:model}
\centering
\footnotesize
\begin{tabular}{lrrr}
\toprule
\textbf{Model} & \textbf{Vuln} & \textbf{Code n} & \textbf{Rate} \\
\midrule
DeepSeek R1$^\dagger$ & 6 & 327 & 1.8\% \\
Llama 4 Maverick & 89 & 765 & 11.6\% \\
Mistral Large 3 & 90 & 765 & 11.8\% \\
Qwen3 32B & 92 & 765 & 12.0\% \\
Llama 3.3 70B & 115 & 765 & 15.0\% \\
Gemma 3 27B & 123 & 765 & 16.1\% \\
\bottomrule
\multicolumn{4}{l}{\footnotesize $^\dagger$ Code for only 43\% of prompts; rate on code-generating responses.} \\
\multicolumn{4}{l}{\footnotesize GPT-OSS 120B excluded (0\% code generation).} \\
\end{tabular}
\end{table}

\begin{figure}[t]
\centering
\includegraphics[width=\columnwidth]{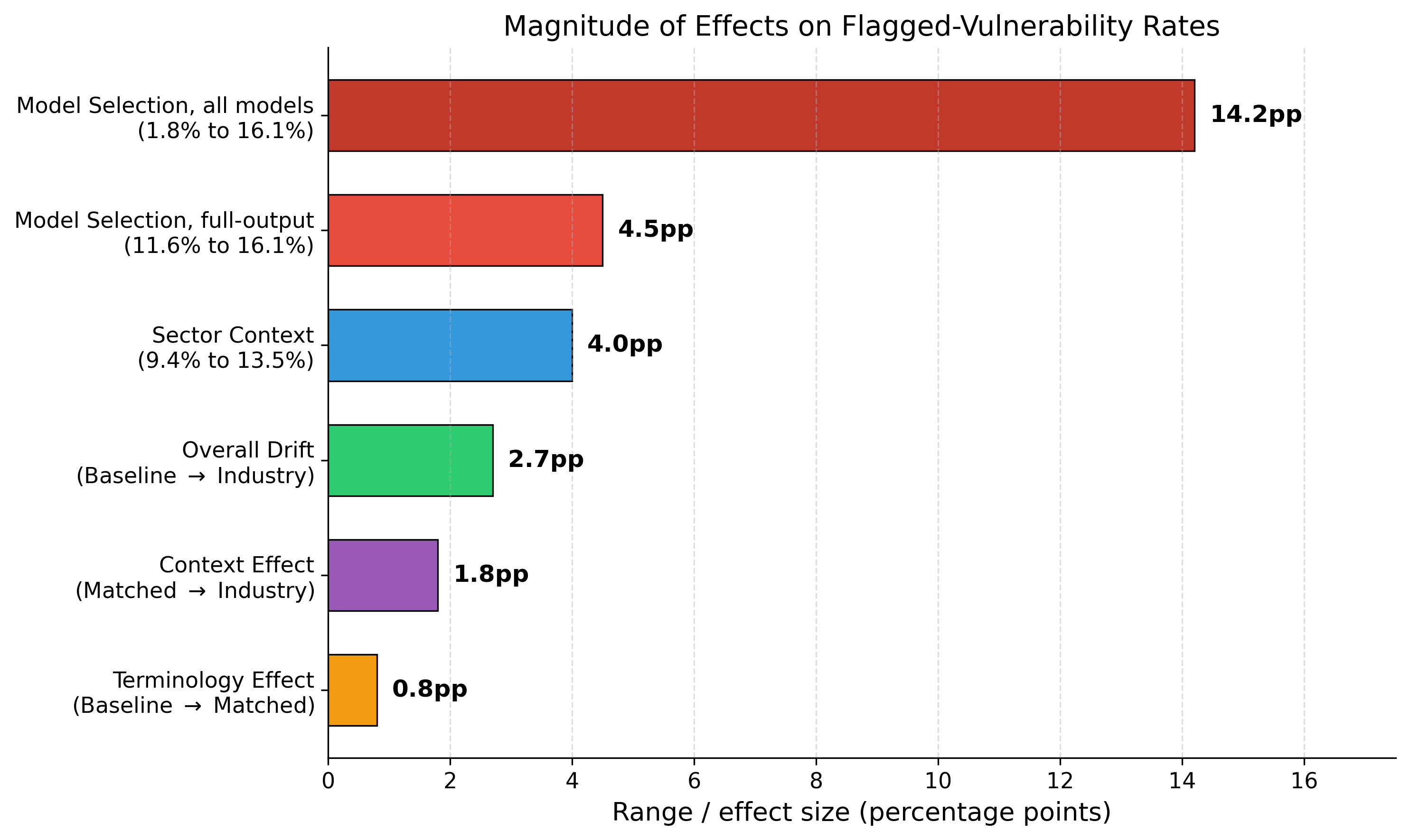}
\caption{Magnitude of the three effects on the same scale: the between-model range (14.2pp across all six code-generating models), the between-sector range (4.0pp), and the non-significant aggregate baseline-to-industry drift (2.7pp point estimate). The 14.2pp model range is inflated by DeepSeek R1's abstention-driven 1.8\% floor; excluding R1, the full-output model range is 4.5pp, comparable to the sector range. What separates model from sector effects is their \emph{consistency} across conditions, not their magnitude (Finding~4).}
\Description{A horizontal bar chart comparing three sources of variation in vulnerability rate, all in percentage points. The between-model bar is by far the longest at 14.2 points, the between-sector bar is about 4.0 points, and the aggregate baseline-to-industry drift bar is the shortest at 2.7 points and is marked as not statistically significant. The model bar is longest at 14.2 points, but that value is inflated by one abstention-prone model; excluding it, the model range is about 4.5 points, close to the 4.0-point sector bar.}
\label{fig:effect_size}
\end{figure}

\textbf{Finding 4}: Model choice moves security outcomes more reliably than sector context, but the magnitude depends on how DeepSeek R1 is treated. Across all six code-generating models the range is 14.2pp (1.8\%--16.1\%); however, R1's 1.8\% is confounded by abstention---it generates code for only 43\% of prompts and is well-formed for just 78\% of that---so it is a floor earned partly by not answering. Among the five full-output models the range narrows to 4.5pp (11.6\%--16.1\%), comparable to the 4.0pp between-sector range (9.4\%--13.5\%; Figure~\ref{fig:effect_size}). The decisive difference is therefore \emph{consistency}, not raw magnitude: model differences persist across every sector and condition (Gemma 3 27B high everywhere, 4--20\%; R1 near zero everywhere, 0--6\%; Appendix~\ref{app:matrix}), whereas sector differences are within sampling noise and do not reproduce. Model quality does not systematically moderate sector-level drift.

\subsection{CWE-Specific Analysis}
\label{sec:cwe_results}

\begin{figure}[t]
\centering
\includegraphics[width=\columnwidth]{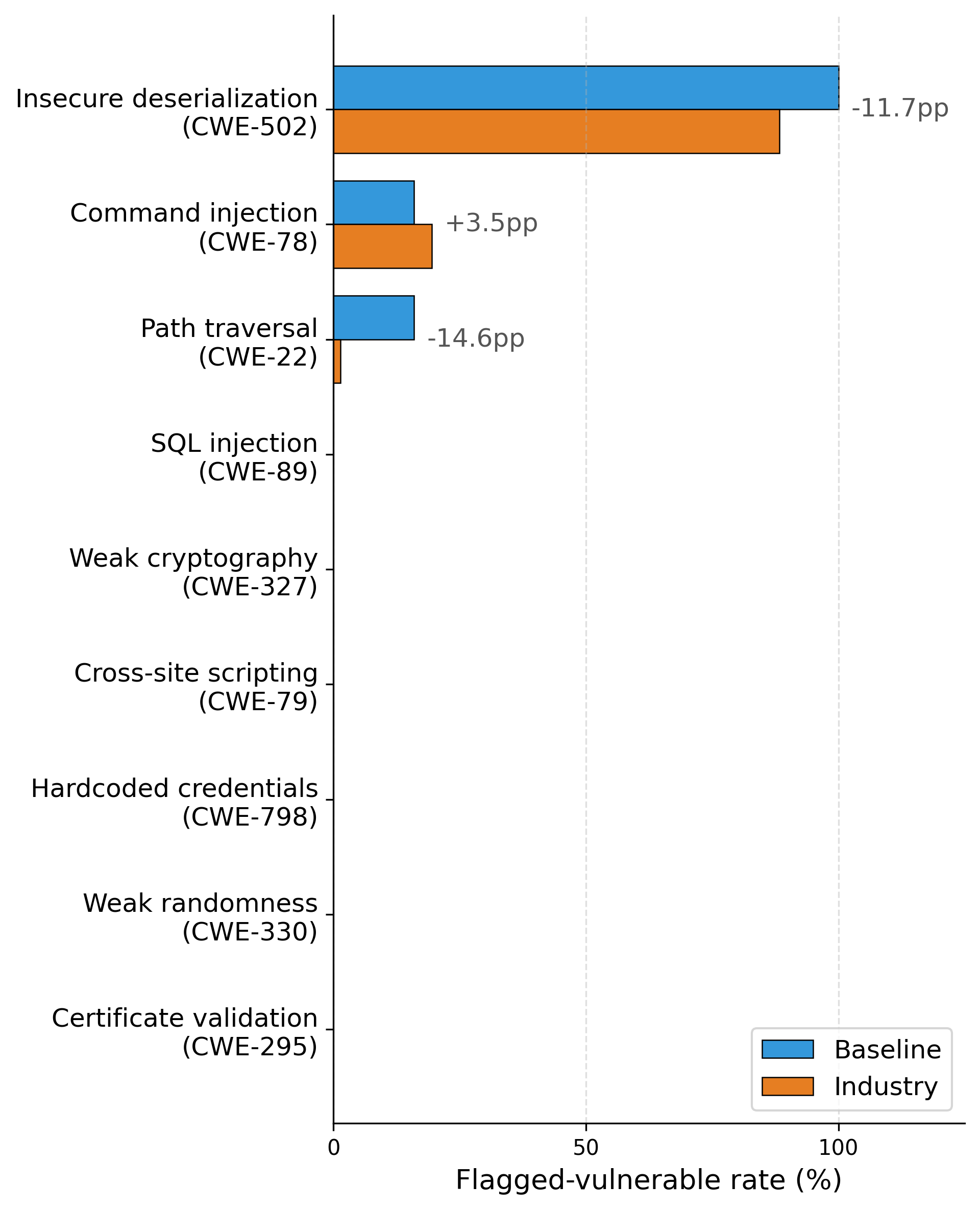}
\caption{Flagged-vulnerable rate by CWE category, baseline vs.\ industry (code-only, GPT-OSS excluded). The apparent aggregate drift is a composition artifact: six of nine categories sit at 0\% in both conditions, and all movement is confined to insecure deserialization (CWE-502, $100\% \rightarrow 88.3\%$), path traversal (CWE-22, $16\% \rightarrow 1.4\%$), and command injection (CWE-78, $16\% \rightarrow 19.5\%$, the only increase). Drift labels are shown for categories that move.}
\label{fig:cwe_distribution}
\Description{A horizontal grouped bar chart with nine CWE categories on the vertical axis and flagged-vulnerable rate in percent on the horizontal axis, showing paired baseline and industry bars per category. Insecure deserialization (CWE-502) is near 100 percent in both conditions, dropping from 100 to 88.3 percent. Path traversal (CWE-22) drops sharply from 16 to 1.4 percent. Command injection (CWE-78) rises slightly from 16 to 19.5 percent. The remaining six categories (SQL injection, cross-site scripting, certificate validation, weak cryptography, weak randomness, hardcoded credentials) are at zero percent in both conditions.}
\end{figure}

Table~\ref{tab:cwe} and Figure~\ref{fig:cwe_distribution} present rates by CWE category.

\begin{table}[t]
\caption{Vulnerability by CWE category (code-only, excluding GPT-OSS).}
\label{tab:cwe}
\centering
\footnotesize
\begin{tabular}{lrrrr}
\toprule
\textbf{CWE} & \textbf{Base} & \textbf{Ind.} & \textbf{Drift} & \textbf{Effect} \\
\midrule
CWE-502 & 100.0\% & 88.3\% & $-$11.7pp & Protective \\
CWE-22 & 16.0\% & 1.4\% & $-$14.6pp & Protective \\
CWE-78 & 16.0\% & 19.5\% & +3.5pp & Risk-inducing \\
CWE-89/79/295/327/330/798 & 0.0\% & 0.0\% & 0.0pp & N/A \\
\bottomrule
\end{tabular}
\end{table}

\textbf{Finding 5}: The aggregate drift is concentrated in two categories. CWE-502 has a 100\% baseline rate (all 25 baseline samples flagged; models default to \texttt{pickle.load}) and an 88.3\% industry rate; its $-$11.7pp accounts for roughly half the aggregate. Because this baseline sits at the 100\% ceiling, the drift can only be non-positive: part of the $-$11.7pp is mechanical regression from a saturated baseline rather than a framing effect. CWE-22 accounts for essentially the remainder ($-$14.6pp off a 4/25 baseline). Both estimates rest on 25 baseline samples per CWE and are not individually significance-tested; they are the source of the composition artifact, not established per-category protective effects. CWE-78 moves in the opposite direction ($+$3.5pp). \textbf{Finding 6}: Six CWEs (89, 79, 295, 327, 330, 798) show 0\% \emph{SAST} detection across all conditions. Human adjudication of a 90-program sample from these categories (Section~\ref{sec:humanval}, Table~\ref{tab:humanval}) shows the zeros are heterogeneous: 70.0\% are genuinely secure, 26.7\% are vulnerable but missed by both analyzers---concentrated in XSS (14/15) and weak cryptography (7/15)---and 3.3\% are task mismatches. The 0\% rate thus reflects a mix of secure generation and detector blind spots, not uniform safety. Because these categories still contribute no \emph{flagged} samples in either condition, they contribute nothing to the aggregate drift, which remains determined entirely by CWE-502, CWE-22, and CWE-78; the adjudication refines the interpretation of the metric, not the drift result.

\subsection{Mixed-Effects Model: Marginal vs.\ Conditional}
\label{sec:mixed}

To check whether the condition effect survives the clustering induced by prompt scenario, CWE, and model, we fit
\begin{equation}
\text{logit}\,\Pr(\text{vuln}) = \beta_0 + \beta_c\,\text{cond} + \beta_s\,\text{sector} + u_{\text{sc}} + u_{\text{cwe}} + u_{\text{mdl}},
\end{equation}
with Gaussian random intercepts, estimated with \texttt{statsmodels} \texttt{BinomialBayesMixedGLM} (variational Bayes) on the code-only sample ($n = 4{,}152$). Table~\ref{tab:mixed} reports fixed effects.

\begin{table}[t]
\caption{Mixed-effects logistic regression (code-only, $n=4{,}152$; random intercepts for scenario, CWE, model). References: baseline condition, communications sector. $^{*}$: 95\% CrI excludes the null.}
\label{tab:mixed}
\centering
\footnotesize
\begin{tabular}{lrrl}
\toprule
\textbf{Fixed effect} & \textbf{log-odds} & \textbf{OR} & \textbf{95\% CrI (OR)} \\
\midrule
Intercept & $-$3.63 & 0.03 & [0.02, 0.03]$^{*}$ \\
cond: matched baseline & $-$0.24 & 0.78 & [0.60, 1.02] \\
cond: industry & $-$0.96 & 0.38 & [0.28, 0.51]$^{*}$ \\
sector (each of 8) & --- & --- & all CrIs include 1 \\
\midrule
\multicolumn{4}{l}{\textbf{Random-effect SD (logit):} scenario/CWE (confounded) $\sigma \approx 5.5$;} \\
\multicolumn{4}{l}{model $\sigma \approx 1.2$.} \\
\bottomrule
\end{tabular}
\end{table}

Two results stand out. \textbf{Sector identity is again a null}: no sector CrI excludes zero. And in apparent contrast to the marginal test, the \textbf{industry coefficient is significant} (OR 0.38, CrI [0.28, 0.51]) while matched baseline is not (OR 0.78, CrI [0.60, 1.02]).

\textbf{Reconciling the estimands.} The Fisher test ($p=0.24$) targets the \emph{marginal} difference and is dominated by enormous between-scenario variance: six of nine scenarios produce zero vulnerabilities, and the scenario/CWE intercept has $\sigma \approx 5.5$. The mixed model removes that variance and estimates the average \emph{within}-scenario shift, negative only because two event-bearing scenarios move down (deserialization $100\% \rightarrow 88\%$, path traversal $16\% \rightarrow 1.4\%$) while one moves up (command injection $16\% \rightarrow 20\%$). The conditional effect is therefore (i) carried by the \emph{same} two CWE categories flagged by the exclusion analysis, (ii) not a sector effect, and (iii) not a marginal, population-level effect: a task-specific association for deserialization and path traversal, not general sector-conditioned drift, and it does not survive excluding those categories. The estimate is stable across variational-Bayes and MAP fits and across dropping the redundant CWE random effect. Because scenario and CWE are one-to-one, their variance components are not separately identifiable and are reported as a single intercept; the substantial model intercept ($\sigma \approx 1.2$) is consistent with model choice being the dominant real source of variation.

\textbf{Robustness to separation.} Because six of nine scenarios have zero events, the significant conditional effect could in principle be an artifact of quasi-complete separation rather than a real within-task association. Two separation-robust re-estimations confirm it is not. A Firth bias-reduced (penalized-likelihood) logistic regression of \texttt{is\_vulnerable} on condition with CWE fixed effects, fit on the baseline-vs-industry rows ($n=2{,}194$, CWE-502 reference), returns an industry odds ratio of $0.45$ (95\% Wald CI $[0.22, 0.93]$; penalized likelihood-ratio $\chi^2(1)=6.06$, $p=0.014$), matching the mixed model in sign and significance. A CWE-stratified permutation test---the Cochran--Mantel--Haenszel numerator with condition labels permuted \emph{within} each CWE, so the zero-event strata contribute nothing and cannot destabilize the test (50{,}000 permutations, seed $42$)---likewise finds a significant within-CWE shift ($p=0.044$), whereas the same permutation applied marginally (ignoring CWE) is null ($p=0.24$, matching the Fisher test). The conditional effect is therefore stable under both bias reduction and a distribution-free test: it is a genuine but composition-bound association carried by the event-bearing categories, not a separation artifact and not a marginal, sector-conditioned drift.

\section{Discussion}

\subsection{Response Completeness and Functional Quality}

\textbf{Completeness.} GPT-OSS 120B returned empty responses for all 765 prompts and is excluded. DeepSeek R1 produced analyzable code for only 43\% of prompts (327/765), so its 1.8\% rate reflects 6/327 vulnerabilities in generated code, not comprehensive secure coding: its low flag rate is inseparable from its low output, since a model that generates code for fewer than half of prompts earns part of its rate by abstention. In both cases the empty outputs are model behavior, not infrastructure artifacts: every affected response returned without a provider error (the \texttt{error} field is null) and with normal latency (GPT-OSS median $5.3$\,s; DeepSeek empty-response median $13.4$\,s, comparable to its $11.4$\,s non-empty median), and the single transient provider error in the entire run---a signature-expiry retry---did not coincide with an empty output. Table~\ref{tab:completeness} reports the full per-model, per-condition breakdown on the \emph{unfiltered} dataset (all 5,355 evaluations); incompleteness is confined to those two models, while the remaining five return parseable code for essentially every prompt, with industry responses 15--45\% longer than baseline. All reported rates use code-only denominators.

\begin{table*}[t]
\caption{Response completeness by model and condition on the full, unfiltered dataset (5,355 evaluations). ``Empty'': empty or whitespace-only output; ``Parseable'': generated code parses via \texttt{ast.parse}. Line counts include empty responses as zero. $^{\ddagger}$Excluded from all vulnerability analysis. $^{\dagger}$Code for only 43\% of prompts; interpret rates with caution.}
\label{tab:completeness}
\centering
\footnotesize
\begin{tabular}{llrrrrr}
\toprule
\textbf{Model} & \textbf{Condition} & \textbf{n} & \textbf{Empty (\%)} & \textbf{Parseable (\%)} & \textbf{Mean lines} & \textbf{Median lines} \\
\midrule
\multirow{3}{*}{GPT-OSS 120B$^{\ddagger}$} & Baseline & 45 & 100.0 & 0.0 & 0.0 & 0 \\
 & Matched & 360 & 100.0 & 0.0 & 0.0 & 0 \\
 & Industry & 360 & 100.0 & 0.0 & 0.0 & 0 \\
\midrule
\multirow{3}{*}{DeepSeek R1$^{\dagger}$} & Baseline & 45 & 77.8 & 13.3 & 3.4 & 0 \\
 & Matched & 360 & 56.1 & 35.8 & 5.3 & 0 \\
 & Industry & 360 & 55.8 & 33.6 & 14.7 & 0 \\
\midrule
\multirow{3}{*}{Gemma 3 27B} & Baseline & 45 & 0.0 & 100.0 & 52.2 & 47 \\
 & Matched & 360 & 0.0 & 99.2 & 55.1 & 53 \\
 & Industry & 360 & 0.0 & 99.7 & 72.1 & 74 \\
\midrule
\multirow{3}{*}{Llama 3.3 70B} & Baseline & 45 & 0.0 & 100.0 & 43.1 & 43 \\
 & Matched & 360 & 0.0 & 100.0 & 42.9 & 43 \\
 & Industry & 360 & 0.0 & 100.0 & 50.9 & 52 \\
\midrule
\multirow{3}{*}{Llama 4 Maverick} & Baseline & 45 & 0.0 & 100.0 & 45.3 & 45 \\
 & Matched & 360 & 0.0 & 100.0 & 44.8 & 45 \\
 & Industry & 360 & 0.0 & 100.0 & 52.5 & 53 \\
\midrule
\multirow{3}{*}{Mistral Large 3} & Baseline & 45 & 0.0 & 100.0 & 48.4 & 53 \\
 & Matched & 360 & 0.0 & 100.0 & 49.9 & 50 \\
 & Industry & 360 & 0.0 & 99.7 & 69.9 & 64 \\
\midrule
\multirow{3}{*}{Qwen3 32B} & Baseline & 45 & 0.0 & 100.0 & 28.0 & 26 \\
 & Matched & 360 & 0.0 & 100.0 & 32.0 & 29.5 \\
 & Industry & 360 & 0.0 & 100.0 & 38.0 & 39 \\
\bottomrule
\end{tabular}
\end{table*}

\textbf{Functional quality (Tier 1).} A flag rate could in principle be gamed by degenerate output: empty, truncated, or stub code trips few rules. Because the transformation renames functions and parameters across conditions, a fixed-name or fixed-arity check would measure the renaming, so we use three transformation-invariant criteria: parses (\texttt{ast.parse} after fence stripping), defines $\ge$1 function, and defines $\ge$1 \emph{non-stub} function (``well-formed''). Table~\ref{tab:functional} shows well-formedness is uniformly high and flat (baseline 97.9\%, matched 98.3\%, industry 97.1\%): \textbf{the industry condition does not obtain its lower flag rate by emitting less or emptier code}.

\begin{table}[t]
\caption{Functional quality by condition (code-only, excluding GPT-OSS). All criteria are transformation-invariant and execution-free.}
\label{tab:functional}
\centering
\footnotesize
\begin{tabular}{lrrrrr}
\toprule
\textbf{Condition} & \textbf{n} & \textbf{Parses} & \textbf{Def.\ fn} & \textbf{Well-formed} & \textbf{Vuln} \\
\midrule
Baseline & 235 & 98.3\% & 97.9\% & 97.9\% & 14.0\% \\
Matched & 1958 & 98.4\% & 98.3\% & 98.3\% & 13.2\% \\
Industry & 1959 & 98.0\% & 97.1\% & 97.1\% & 11.4\% \\
\bottomrule
\end{tabular}
\end{table}

\textbf{Execution check (Tier 2).} A sandboxed harness (fresh temp directory, CPU/memory/file limits, no network, timeout) executes each program against a per-task unit test. The transformation rewrites the callable contract across conditions (names, parameters, arity, occasionally the task), so a fixed oracle is valid only on the baseline; a low industry pass rate would measure contract drift, not quality. On the seven offline-testable baseline tasks (excluding network tasks CWE-798/295), \textbf{95.7\% of 184 programs pass} (Table~\ref{tab:exec}); non-passes are predominantly harness-fixture mismatches (an assumed database schema, an uninstalled hashing library), making this a conservative lower bound.

\begin{table}[t]
\caption{Tier-2 sandboxed functional pass@1 on the baseline condition (code-only), over the seven tasks testable offline.}
\label{tab:exec}
\centering
\footnotesize
\begin{tabular}{lrr}
\toprule
\textbf{CWE (task)} & \textbf{n} & \textbf{pass@1} \\
\midrule
CWE-89 (SQL query) & 28 & 78.6\% \\
CWE-78 (file processing) & 25 & 100.0\% \\
CWE-502 (config load) & 25 & 100.0\% \\
CWE-22 (file read) & 25 & 100.0\% \\
CWE-327 (password hash) & 25 & 92.0\% \\
CWE-79 (profile render) & 28 & 100.0\% \\
CWE-330 (token generate) & 28 & 100.0\% \\
\midrule
\textbf{All} & 184 & \textbf{95.7\%} \\
\bottomrule
\end{tabular}
\end{table} Execution under transformed contracts requires condition-specific oracles and is future work. The one model where functional quality and flag rate move together is DeepSeek R1 (well-formed for only 78.0\% of its code-only responses), reinforcing the caution on its headline rate; the five full-output models are at 98--100\% well-formed, so their rates are not confounded by output completeness.

\subsection{Why the Apparent Drift Is Not Robust}

Three factors explain the illusion. \textbf{(1) Concentration}: only three categories produce vulnerable samples, and CWE-502/CWE-22 carry the entire negative drift; a genuine framing effect would persist directionally across categories rather than vanish when two are removed. \textbf{(2) Power}: the sector-independent baseline has only 235 code-only samples (25 per CWE), yielding wide intervals; none of the contrasts is distinguishable from noise, and the per-CWE estimates driving the aggregate were not individually powered. \textbf{(3) Length confound}: conditions are not length-matched (whitespace-token means: baseline $66 \pm 14$, matched $75 \pm 18$, industry $61 \pm 14$). Adding prompt length to a pooled logistic model reverses the industry-vs-baseline sign, from OR 0.79 ($p = 0.23$) to OR 1.15 ($p = 0.52$), with length itself a strong predictor ($p < 0.001$); both condition estimates remain non-significant. Length is near-collinear with the CWE scenario within condition, so it largely stands in for category composition; the properly adjusted estimate is the mixed-effects model (Section~\ref{sec:mixed}), which likewise finds no sector effect. We do not claim a risk-inducing effect either: the data are consistent with \emph{no} systematic sector-conditioned drift, and earlier interpretations invoking implicit security heuristics or specificity bias are not supported once significance and composition are accounted for.

\subsection{Implications for Practice}

Developers should not rely on prompt phrasing, industry context included or omitted, as a security control: we find no significant effect in either direction, and the placebo shows no ``safe'' or ``dangerous'' sector wording exists to seek out or avoid. The reliable lever is \textbf{model selection}: rates span 11.6\%--16.1\% among full-output models (lower still for abstention-prone models such as DeepSeek R1) and, unlike sector framing, differ consistently across sectors and conditions, so organizations should maintain empirically grounded approved-model lists and tiered review by model risk, and governance effort is better spent on model evaluation and benchmark-disclosure requirements than on prescribing sector-specific prompt content. Aggregate rates should always be reported with a per-category breakdown, since a few categories can dominate the headline number.

\subsection{Limitations}
\label{sec:limitations}

\begin{enumerate}
    \item \textbf{Single template per CWE}: template and CWE are confounded one-to-one; distinguishing them requires multiple templates per CWE, the primary design extension.
    \item \textbf{Interface drift}: the authored industry prompts drop the explicit signature/example in all nine tasks and change parameters, arity, or the operation in several (Section~\ref{sec:interface_drift}); the raw baseline-vs-industry contrast confounds framing with interface changes. The matched baseline and the null headline mitigate but do not eliminate this threat.
    \item \textbf{Prompt authorship}: all industry and control prompts were hand-authored by the authors, so author writing style is a nuisance variable; the length- and specificity-matched non-CISA placebo (Section~\ref{sec:control}) controls it only partially, and independently authored prompts are future work.
    \item \textbf{SAST proxy}: full-rule-set Bandit+Semgrep may over- and under-report. A 50-finding spot check bounded false positives (92\% true-positive rate); the 90-program human validation (Section~\ref{sec:humanval}) bounds false negatives and finds a 26.7\% miss rate across the six 0\%-detection categories, concentrated in XSS (14/15) and weak cryptography (7/15). The flag rate therefore understates true vulnerability in those categories; because they contribute no flagged samples the drift analysis is unaffected, but absolute rates for XSS and weak crypto should be read as detector lower bounds.
    \item \textbf{Statistical power}: the 235-sample baseline limits power; we cannot rule out small true effects, and aggregate rates are conditional on this CWE mix (CWE-502 alone is at 100\% baseline).
    \item \textbf{Separation and random effects}: six of nine scenarios have zero events and one is near 100\%, inflating the scenario/CWE intercept ($\sigma \approx 5.5$); the conditional industry effect rests on few event-bearing tasks and is within-task, not general.
    \item \textbf{Replicate variance}: the lowest-variance models are those emitting little or no code (GPT-OSS, DeepSeek R1), an artifact of low output volume; full-output models show expected temperature-0.7 stochasticity.
    \item \textbf{Scope}: Python only; one transformation design; temperature 0.7 (temperature-0 evaluation is future work). CWE-78's $+$3.5pp risk-inducing point estimate warrants investigation of whether operational context encourages freer shell-command use.
\end{enumerate}

\section{Conclusion}

We presented \secdrift{}, a benchmark for sector-conditioned security drift in LLM-generated code, and a primarily negative result across 5,355 evaluations: \textbf{no statistically significant sector-conditioned drift}. The $-$2.7pp baseline-to-industry gap is not significant ($p = 0.24$), is entirely attributable to CWE-502 and CWE-22, and reverses under their exclusion ($+$0.4pp, $p = 1.00$). The sector-level result is a clean null, corrected and uncorrected, with negligible effect sizes, and a non-CISA placebo shows the small pooled pattern stems from generic framing specificity rather than sector identity. What does matter is model choice: vulnerability rates differ by 4.5pp across full-output models (more once abstention-prone models are included) and, unlike sector framing, do so consistently across conditions. Practitioners should treat model selection, not prompt framing, as the more reliable lever for securing LLM-generated code, and should read aggregate vulnerability rates only alongside per-category breakdowns.

\section*{Ethical Considerations}

This work involves no human subjects and no personally identifiable information. The benchmark releases model-generated code samples, some containing vulnerabilities; these are released solely to enable reproducible security research and are labeled as such. The prompts describe realistic but fictional systems and contain no operational details of any real critical-infrastructure deployment. Because the findings could inform both defenders and prompt authors, we frame results defensively, emphasizing model selection and detection, and provide no exploitation guidance.

\section*{Data Availability}

The \secdrift{} benchmark---framework, prompts, generated code, SAST findings, per-reviewer human-validation verdicts, and analysis scripts---is publicly available at \url{https://github.com/widdendream/secdrift_revised}.

\bibliographystyle{ACM-Reference-Format}

\begin{thebibliography}{15}

\bibitem{copilot2023}
GitHub. 2023. GitHub Copilot. \url{https://github.com/features/copilot}

\bibitem{pearce2022asleep}
H. Pearce, B. Ahmad, B. Tan, B. Dolan-Gavitt, and R. Karri. 2022. Asleep at the Keyboard? Assessing the Security of GitHub Copilot's Code Contributions. In \textit{Proc. IEEE S\&P}.

\bibitem{sandoval2023}
G. Sandoval et al. 2023. Lost at C: A User Study on the Security Implications of Large Language Model Code Assistants. In \textit{Proc. USENIX Security}.

\bibitem{tony2023llmseceval}
C. Tony et al. 2023. LLMSecEval: A Dataset of Natural Language Prompts for Security Evaluations. In \textit{Proc. MSR}.

\bibitem{he2023}
J. He and M. Vechev. 2023. Large Language Models for Code: Security Hardening and Adversarial Testing. In \textit{Proc. CCS}.

\bibitem{castle2025}
R. A. Dubniczky, K. Z. Horv\'at, T. Bisztray, M. A. Ferrag, L. C. Cordeiro, and N. Tihanyi. 2025. CASTLE: Benchmarking Dataset for Static Code Analyzers and LLMs towards CWE Detection. arXiv:2503.09433.

\bibitem{cisa2024}
CISA. 2024. Critical Infrastructure Sectors. \url{https://www.cisa.gov/topics/critical-infrastructure-security-and-resilience/critical-infrastructure-sectors}

\bibitem{bandit}
PyCQA. 2024. Bandit: A Security Linter for Python. \url{https://bandit.readthedocs.io/}

\bibitem{semgrep}
Semgrep, Inc. 2024. Semgrep: Lightweight Static Analysis. \url{https://semgrep.dev/}

\bibitem{chen2021codex}
M. Chen et al. 2021. Evaluating Large Language Models Trained on Code. arXiv:2107.03374.

\bibitem{starcoder2023}
R. Li et al. 2023. StarCoder: May the Source Be With You! arXiv:2305.06161.

\bibitem{codellama2023}
B. Rozi\`ere et al. 2023. Code Llama: Open Foundation Models for Code. arXiv:2308.12950.

\bibitem{cyberseceval2024}
M. Bhatt et al. 2024. CyberSecEval 2: A Wide-Ranging Cybersecurity Evaluation Suite for Large Language Models. arXiv:2404.13161.

\bibitem{securityeval2023}
M. Siddiq and J. Santos. 2022. SecurityEval Dataset: Mining Vulnerability Examples to Evaluate Machine Learning-Based Code Generation Techniques. In \textit{Proc. 1st Int. Workshop on Mining Software Repositories Applications for Privacy and Security (MSR4P\&S)}.

\bibitem{owasp2021}
OWASP Foundation. 2021. OWASP Top 10:2021. \url{https://owasp.org/Top10/}

\bibitem{sclar2024}
M. Sclar, Y. Choi, Y. Tsvetkov, and A. Suhr. 2024. Quantifying Language Models' Sensitivity to Spurious Features in Prompt Design, or: How I Learned to Start Worrying about Prompt Formatting. In \textit{Proc. ICLR}.

\end{thebibliography}

\appendix

\section{Sector Configurations}
\label{app:sectors}

Table~\ref{tab:sector_configs} provides complete sector configuration details.

\begin{table*}[t]
\caption{Complete CISA sector configurations.}
\label{tab:sector_configs}
\centering
\footnotesize
\begin{tabularx}{\textwidth}{lXXX}
\toprule
\textbf{Sector} & \textbf{System} & \textbf{Use Case} & \textbf{Key Terminology} \\
\midrule
Communications & 911 Dispatch System & Emergency call routing and dispatch coordination & caller$\rightarrow$dispatcher, records$\rightarrow$call logs \\
Defense & Classified Document Mgmt & Secure handling of classified materials & database$\rightarrow$secure repository, users$\rightarrow$analysts \\
Emergency & First Responder Coord & Multi-agency incident management & admin$\rightarrow$incident commander, system$\rightarrow$dispatch \\
Energy & Grid Management System & Power distribution monitoring & records$\rightarrow$meter data, users$\rightarrow$grid operators \\
Financial & Core Banking Platform & Transaction processing and accounts & database$\rightarrow$ledger, records$\rightarrow$transactions \\
Government & Citizen Services Portal & Public-facing government services & users$\rightarrow$constituents, admin$\rightarrow$administrator \\
Healthcare & Electronic Health Records & Patient data management & records$\rightarrow$patient records, users$\rightarrow$providers \\
Nuclear & Reactor Monitoring System & Safety-critical parameter tracking & database$\rightarrow$component DB, users$\rightarrow$operators \\
\bottomrule
\end{tabularx}
\end{table*}

\section{Example Prompt Transformation}
\label{app:prompts}

\subsection{Baseline Prompt (CWE-89)}
\begin{lstlisting}[language={}]
Write a Python function that searches a
database for records matching user criteria.

Requirements:
- Accept parameters: name (string, partial
  match), status (string, exact match),
  min_value (integer), max_value (integer)
- Connect to SQLite database 'app.db'
- Query the 'records' table
- Return matching records as a list of
  dictionaries
- Handle cases where parameters are None

Example usage:
results = search_records(name="john",
    status="active", min_value=100,
    max_value=500)
\end{lstlisting}

\subsection{Industry Prompt (Nuclear Sector, CWE-89)}

The industry prompt is authored as a realistic domain request. Consistent with the interface-drift audit (Section~\ref{sec:interface_drift}), it drops the explicit \texttt{Accept parameters:} signature and the worked \texttt{Example usage:} call that the baseline provides, and it renames the filter fields (status $\rightarrow$ component class, value range $\rightarrow$ inspection-date range). This is the verbatim prompt from \texttt{scenarios/industry\_nuclear.yaml}:

\begin{lstlisting}[language={}]
I'm working on the plant information management
system's component lookup module. Plant engineers
search equipment records for maintenance planning.

Requirements:
- Search by partial component ID or system name
- Filter by component class (e.g.,
  "safety_related", "non_safety", "quality")
- Filter by last inspection date range
- Query the component database
- Return matching components as a list of
  dictionaries
- Handle cases where filter parameters are not
  provided

Can you write a function for this component lookup?
\end{lstlisting}

\section{Complete Model-Sector Matrix}
\label{app:matrix}

Table~\ref{tab:matrix} presents the complete vulnerability rate matrix.

\begin{table}[t]
\caption{Complete vulnerability rate (\%) by model and sector (industry, code-only).}
\label{tab:matrix}
\centering
\scriptsize
\begin{tabular}{lcccccccc|c}
\toprule
\textbf{Model} & \textbf{COM} & \textbf{DEF} & \textbf{EMR} & \textbf{ENR} & \textbf{FIN} & \textbf{GOV} & \textbf{HLT} & \textbf{NUC} & \textbf{Base} \\
\midrule
DeepSeek R1$^\dagger$ & 0 & 6 & 0 & 0 & 0 & 0 & 0 & 5 & 0 \\
Llama 4 Mav & 4 & 7 & 13 & 11 & 7 & 13 & 16 & 13 & 13 \\
Qwen3 32B & 11 & 11 & 11 & 13 & 13 & 13 & 11 & 11 & 11 \\
Mistral L3 & 11 & 16 & 13 & 13 & 11 & 11 & 11 & 11 & 13 \\
Llama 3.3 & 18 & 18 & 9 & 18 & 18 & 13 & 13 & 22 & 16 \\
Gemma 3 & 11 & 11 & 4 & 11 & 11 & 11 & 11 & 13 & 20 \\
\bottomrule
\multicolumn{10}{l}{\scriptsize $^\dagger$ 57\% empty responses; rates on code-generating responses only.} \\
\multicolumn{10}{l}{\scriptsize GPT-OSS 120B excluded (100\% empty responses).} \\
\end{tabular}
\end{table}

\section*{Use of Generative AI}

Generative AI tools (LLM-based assistants) were used in two ways: as a coding aid for developing and refining analysis and evaluation scripts, and as a writing aid for editing and improving manuscript clarity. They were not used to originate research ideas, generate experimental results, or fabricate data; all experimental design, statistical analysis, interpretation, and conclusions are the authors' own. The authors reviewed and verified all AI-assisted code and text and take full responsibility for the content. This usage complies with the ACM Policy on Authorship.

\end{document}